\documentclass[reprint,superscriptaddress,amsmath,amssymb,aip,showkeys]{revtex4-2}

\usepackage{graphicx}
\usepackage{dcolumn}
\usepackage{xcolor}

\usepackage{enumitem}
\setlist{itemsep = 0.0pt}

\usepackage{cancel}

\usepackage{booktabs}

\definecolor{nblue}{RGB}{0, 0, 0}
\newcommand{\blue}[1]{\textcolor{nblue}{#1}}

\begin{document}

\title{Computational design of moir{\'e} assemblies aided by artificial intelligence}

\author{Georgios A. Tritsaris}
\affiliation{John A. Paulson School of Engineering and Applied Sciences, Harvard University, Cambridge, Massachusetts 02138, USA}
\author{Stephen Carr}
\affiliation{Brown Theoretical Physics Center and Department of Physics, Brown University, Providence, Rhode Island 02912, USA}
\author{Gabriel R. Schleder}
\affiliation{John A. Paulson School of Engineering and Applied Sciences, Harvard University, Cambridge, Massachusetts 02138, USA}
\affiliation{Federal University of ABC (UFABC), Santo Andr{\'e}, S{\~a}o Paulo 09210-580, Brazil}

\date{\today}

\begin{abstract}
Two-dimensional (2D) layered materials, demonstrating significantly different properties from their bulk counterparts, offer a materials platform with potential applications from energy to information processing devices. Although some single- and few-layer forms of materials such as graphene and transition metal dichalcogenides have been realized and thoroughly studied, the space of arbitrarily layered assemblies is still mostly unexplored. \blue{The main goal of this work is to demonstrate precise control of layered materials' electronic properties through careful choice of the constituent layers, their stacking, and relative orientation. Physics-based and AI-driven approaches for the automated planning, execution, and analysis of electronic structure calculations are applied to layered assemblies based on prototype one-dimensional (1D) materials and realistic 2D materials. We find it is possible to routinely generate moir{\'e} band structures in 1D with desired electronic characteristics such as a band gap of any value within a large range, even with few layers and materials (here, four and six, respectively). We argue that this tunability extends to 2D materials by showing the essential physical ingredients are already evident in calculations of two-layer MoS$_2$ and multi-layer graphene moir{\'e} assemblies.}
\end{abstract}


\maketitle

\section{Introduction} \label{sec:introduction}
Fostering innovation in nanotechnology relies on the continuous development of nanostructures with exceptional properties as active materials. Graphene and other two-dimensional (2D) materials such as transition metal dichalcogenides constitute structurally simple, but nevertheless fascinating, examples of materials that could redefine information processing, communication, energy storage, and a host of other technologies \citep{Geim2013Van,novoselov_2d_2016}. In particular, precise and rapid stacking of single-layer units from an ever-expanding library of layered materials to form 2D layered assemblies with target properties is expected to significantly reduce barriers to device development and commercialization \citep{Zhou2018library,masubuchi_autonomous_2018}. 

Manipulating the properties of 2D layered assemblies through their twist angle has emerged as a new paradigm in materials design \citep{Carr2017Twistronics,tritsaris_electronic_2020,Bistritzer2011Moire}. Recently, Cao {\em et al.} \citep{Cao2018Unconventional,Cao2018Correlated} reported the experimental observation that when two sheets of graphene are stacked together and twisted at a small, so-called magic, angle, the resulting superlattice can become either an insulator or a superconductor. Layered assemblies of graphene, hexagonal boron nitride (hBN), molybdenum disulfide (MoS$_2$) and other 2D materials are now routinely fabricated in the lab \citep{furchi_photovoltaic_2014,loan_graphenemos2_2014,lee_flexible_2013}. Notably, the work of Masubuchi {\em et al.} \citep{masubuchi_autonomous_2018} introduced an approach to the high-throughput robotic assembly of 2D crystals for designer multi-layer moir{\'e} superlattices. 

\begin{figure}
  \centering
  \includegraphics[width=\columnwidth]{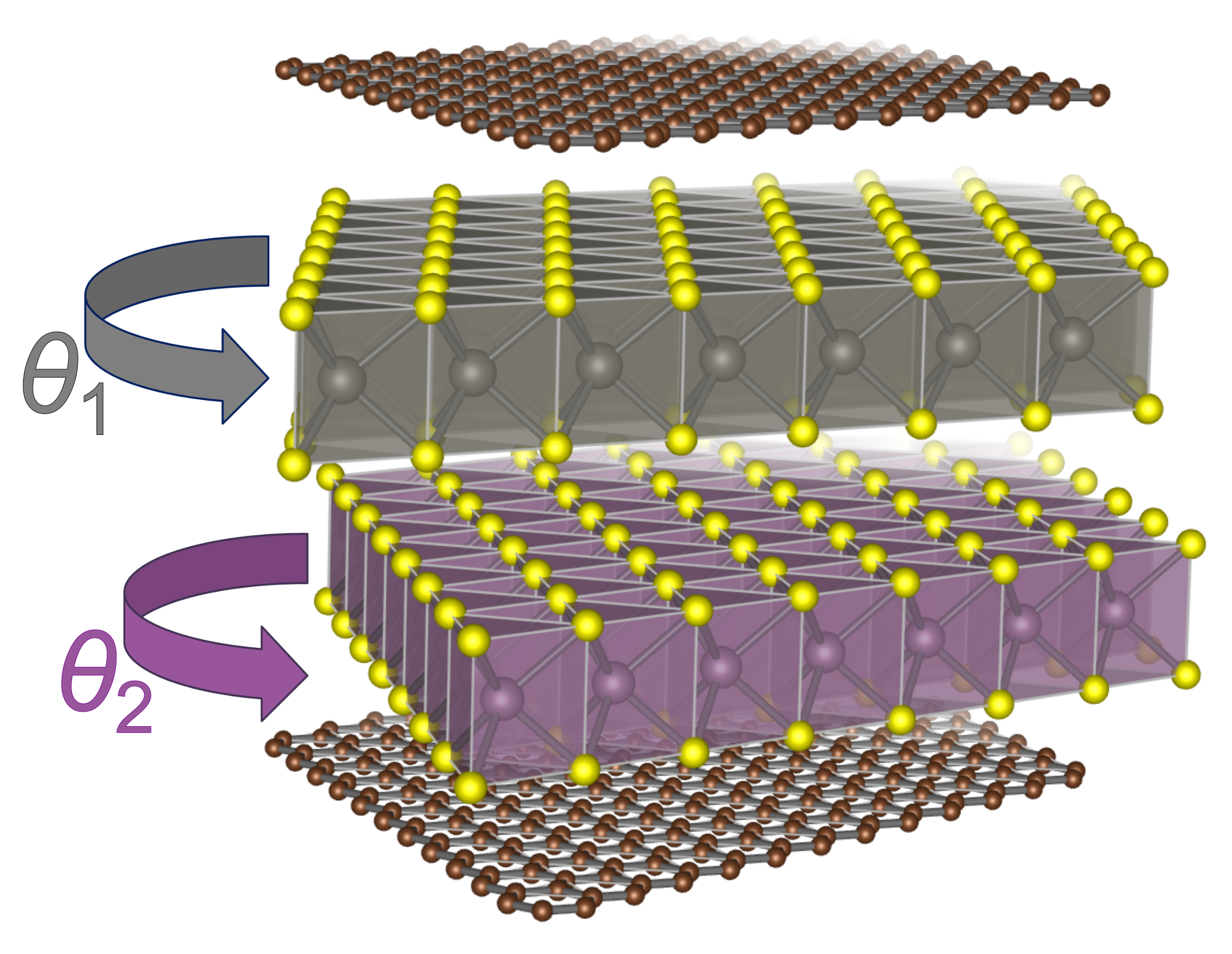}
\caption{a) Atomistic representation of a model layered assembly (counterclockwise twist angles $\theta$).}
\label{fig:structure}
\end{figure}

However, even with a small number of materials (or, within the context of 2D layered assemblies, atomically-thin single-layer units or building blocks; Figure~\ref{fig:structure}), screening the resulting configuration space for structures with potentially interesting physical behavior becomes time-consuming and resource-demanding, if not an unattainable goal, due to the combinatorial nature of the design problem, {\em i.e.,} the exceedingly large number of possible stacking sequences and layer orientations. Not only do electronic transport measurements depend sensitively on the twist angle, but typically only a small number of materials can be prepared and measured over relevant time frames due to the delicate nature of their fabrication.

In contrast, high-throughput materials calculations introduce the capability to investigate large libraries of materials, both for elucidating the physical processes that govern their electronic, optical, magnetic, and other properties, and for the discovery of novel structures \citep{bassman_active_2018,Haastrup2018Computational,tritsaris_electronic_2020,choudhary_high-throughput_2017,schleder_exploring_2020,rhone_data-driven_2020}. The {\em in silico} screening and design of 2D layered materials has become an increasingly attractive option, owing in no small part to the continuing development of novel theoretical concepts and methods for the study of their atomic and electronic structure \citep{Tritsaris2016Perturbation,Carr2017Twistronics,thygesen_calculating_2017}. For instance, theoretical treatment of layered assemblies of graphene has established features such as low-dispersion (or flat) electronic bands in the single-particle band structure as indicators of interesting electronic behavior \citep{Bistritzer2011Moire,Carr2017Twistronics}. A concise review of electronic structure methods based on density functional theory (DFT), tight-binding (TB) Hamiltonians, and continuum models for moir{\'e} superlattices is provided in Carr {\em et al.} \citep{Carr2020review}

\blue{However, to the best of our knowledge, there is still no universal approach suitable for the calculation of arbitrarily stacked layered materials. Thus, within the context of selection and design of materials, an important question remains: ``{\em is it possible to design a layered assembly that could exhibit any desirable electronic properties?}'' Although the assumption of electronic tunability permeates much of current research in the field of layered materials, we know no examples of specific conclusive evidence. The main goal of this work is then to demonstrate precise control of the electronic properties of layered materials through careful choice of the constituent layers, their stacking, and relative orientation.}

Recent advances at the forefront of machine learning and automation, combined with the continuing expansion of computing resources, have paved the way for a paradigm shift in materials research and development towards seamlessly closing the loop between hypothesis formulation and evaluation. For instance, Kusne {\em et al.} \citep{kusne_title_nodate} relied on a closed-loop, active learning-driven autonomous system for the investigation and discovery of phase-change materials; Attia {\em et al.} \citep{attia_closed-loop_2020} combined a predictive model and a Bayesian optimization algorithm to efficiently optimize fast-charging protocols for maximizing battery cycle life; Burger {\em et al.} \citep{burger_mobile_2020} used a mobile robot to search for improved photocatalysts for hydrogen production; and the work of Montoya {\em et al.} \citep{montoya_autonomous_2020} presented a computational system for optimization in large search spaces of materials by adopting an agent-based approach to deciding which experiments to carry out. \blue{Here, we use agent-based simulation to explore and broaden the space of layered materials.} 

More specifically, we introduce physics-based and AI-driven approaches for the automated planning, execution and analysis of (virtual) materials measurements, \blue{and apply them to investigate trends in the electronic structure of layered assemblies based on prototype one-dimensional (1D) materials, motivated by realistic 2D materials. We find that these prototype heterostructures offer a wide range of electronic band structure properties, with indirect band gaps covering the entire spectrum between $0$ to $2.0$ eV, and moir\'e miniband bandwidths and gaps spanning more than three orders of magnitude. This result stems from two important properties of layered materials: (1) each unique sequence of layers generates a unique band gap and interlayer moir\'e potential, and (2) these potentials are smoothly tunable by changing the moir\'e length. These two necessary points regarding electronic tunability are then validated in candidate 2D materials, using calculations of bilayer MoS$_2$ to show smooth twist angle dependence of electronic properties, and multi-layer graphene moir{\'e} assemblies as evidence of the band diversity caused by different numbers and locations of twisted interfaces.}

The manuscript is organized as follows: Section~\ref{sec:concepts} introduces theoretical methods for the calculation of the electronic structure of low-dimensional moir{\'e} assemblies, and a computational framework for automated materials discovery and design using agent-based simulation. Computational, algorithmic, and implementation details are provided therein. \blue{In Section~\ref{sec:results} we demonstrate it is possible to routinely generate moir{\'e} band structures with desired characteristics within a multi-layer and multi-material space of moir{\'e} assemblies with the aid of model 1D materials.} The feasibility of applying these obtained insights to realistic 2D systems is then explored by calculations of two-layer MoS$_2$ and multi-layer graphene moir{\'e} assemblies. Finally, Section~\ref{sec:conclusions} examines the implications of the current work and proposes directions for future research.

\section{Concepts, models, and methods} \label{sec:concepts}
In this section, we introduce concepts, models, and methods for the systematic investigation of arbitrarily layered assemblies.

\subsection{Theoretical framework} \label{sec:theoretical}
Many experimentally relevant materials properties can be derived from single-particle band structures, and as such calculations of electronic bands will be the focus of this work. All electronic structure calculations were performed using TB models of moir{\'e} assemblies, owing to their favorable balance between explanatory power, accuracy, and computational complexity.

To illustrate the central idea of our TB approach for low-dimensional moir{\'e} assemblies in a simple context, we introduce a general 1D chain model \citep{Carr2020duality}. This is described by an empirical interlayer coupling functional that allows the study of how different material combinations control the electronic properties of moir{\'e} band structures. The interlayer couplings, $t_{ij}$, between orbitals $i$ and $j$ of neighboring 1D layers are given by
\begin{equation}
    t_{ij} = \nu e^{- \left(|\Delta {r}_{ij}|/R_0 \right)^2}
\end{equation}
where $\Delta {r}_{ij}$ is the relative distance between the orbitals, $\nu = 1$~eV and $R_0~=~2.5$~\AA{}.
\blue{These values are selected to mimic primarily the strength and length-scale for electronic interlayer coupling in the transition metal dichalcogenides, a popular family of 2D semiconductors \cite{Fang2015Ab}.}
The size of the corresponding Hamiltonians increases at an order of magnitude less than those required for 2D systems.

\begin{figure*}
  \centering
  \includegraphics[width=\textwidth]{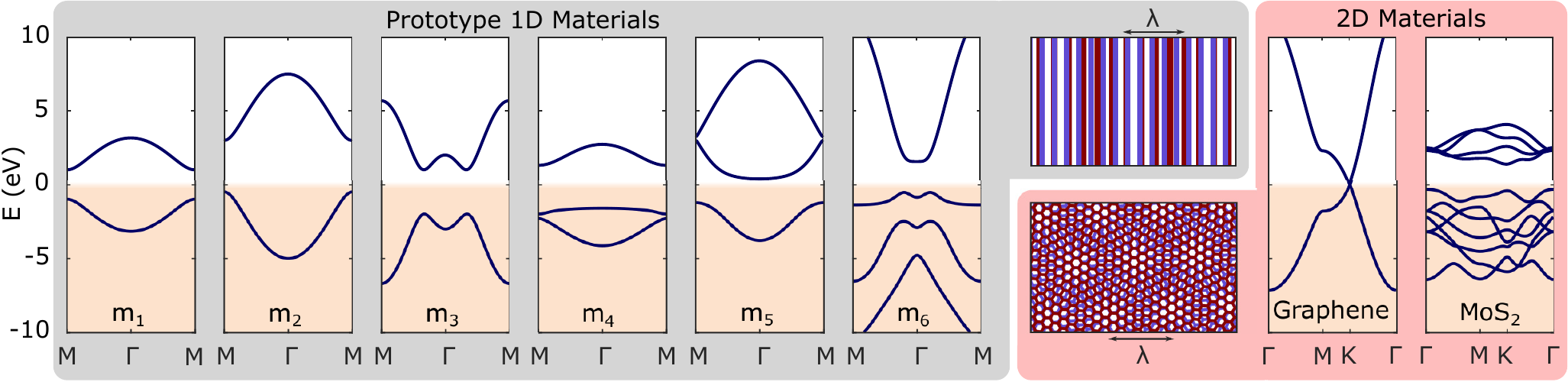}
\caption{Tight-binding band structures of monolayers for (left) six prototype 1D semiconductors (symbols m$_1$-m$_6$) and (right) the 2D materials graphene and MoS$_2$. (center) Example 1D and 2D moir{\'e} patterns with moir{\'e} length $\lambda$.}
\label{fig:1Dstructures}
\end{figure*}

It is possible to design moir{\'e} assemblies based on 1D chains with band structures that mimic what is seen in twisted 2D layered assemblies by deliberately choosing their TB coefficients. Figure~\ref{fig:1Dstructures} shows the band structures of six such prototype 1D semiconductors. Their TB Hamiltonians have slightly different band energies for the valence and conduction edges, as well as different band gaps. Without loss of generality, three (m$_1$, m$_2$, and m$_3$ in Figure~\ref{fig:1Dstructures}) are analogous to hBN, a common 2D insulator, having two orbitals per unit cell but differing on-site energies. The sublattice symmetry-breaking opens a gap at the $k = \pi/2$ point. The other three (m$_4$, m$_5$, and m$_6$), with more than two orbitals per cell, have more complicated band structures and were constructed to mimic 2D semiconductors like the transition metal dichalcogenides.

\blue{For the 1D systems, the moir{\'e} patterns are generated using lattice mismatch as an analogue of twist angle. The central idea of the 2D moir\'e patterns is that twisting causes a new effective periodicity $\lambda$ to appear in the multilayer structure. An identical procedure can be implemented in 1D by varying the lattice constants, similarly to the study of moir\'e patterns of graphene on aligned hBN~\cite{Hunt2013,Dean2013}.} The lattice mismatch is described using variable $\Theta$, where one set of layers is uncompressed with unit ($1$) lattice constant, and the other set are compressed with lattice constant of $1 - \Theta$. To ensure periodicity of the superlattices, we used the generating function 
$$\Theta_N = \frac{1}{2(1 + N)},$$
with $N$ an integer in the interval $1 \leq N \leq 9$. These values for $N$ allow for a wide range of moir{\'e} pattern sizes without compromising computational speed by ensuring the TB model of the largest 1D superlattice considered in this work does not exceed 400 orbitals. 

In contrast to the 1D materials, the calculation of the 2D systems, MoS$_2$ and graphene, rely on the Wannier transformation of electronic states obtained from DFT calculations \citep{kohn_self-consistent_1965,marzari_maximally_2012}. By sampling multiple stacking configurations between two single layers of the same 2D material, the twisted system's TB Hamiltonian can be generated with accurate intra- and interlayer couplings. Additional details are provided in the works of Fang {\em et al.} \citep{Fang2015Ab,Fang2016Electronic} 

Since arbitrary rotations generally involve incommensurate superlattices for which a periodic atomistic representation (as would be required to avoid spurious effects from the edges of finite structural models such as flakes) does not exist, we calculated only commensurate superlattices. 2D superlattices were identified based on the formalism of Uchida {\em et al.} \citep{Uchida2014Atomic} for the twist angle $\theta$, in radians, 
$$\theta_N = {\rm cos}^{-1} \left( \frac{c(N)+1}{c(N)+2} \right),$$
where $c(N) = 6N(N+1)$. Specifically, we investigate twisted layered assemblies in the interval $1 \leq N \leq 32$, which corresponds to a set of thirty-one (32) discrete values for the twist angle in the range $1.02^\circ \leq \theta < 21.79^\circ$.

The values of $\Theta$ for the 1D superlattices are comparable to those of the twist angle $\theta$, for the 2D layered structures: for example, $\Theta = 0.05$ (compressive strain, or lattice mismatch, of 5\%) corresponds to $\theta = 0.05 (180^\circ / \pi) = 2.8^\circ$. Although the calculation of twisted layered assemblies in 2D relies on DFT inputs to describe materials realistically, important features in their electronic structure can be understood with fewer complications using 1D model structures. Moreover, these structures are directly comparable to striped moir{\'e} patterns in 2D interfaces, which can occur during the experimental fabrication of a twisted interface \cite{Alden2013}. A detailed discussion provided in Carr {\em et al.} \citep{Carr2020duality} posits an intimate connection between the 1D and 2D moir{\'e} assemblies, justifying the use of these 1D models in the exploration of trends in the electronic structure of twisted 2D materials.

\subsection{Computational framework} \label{sec:computational}

\begin{figure*}
  \centering
  \includegraphics[width=\textwidth]{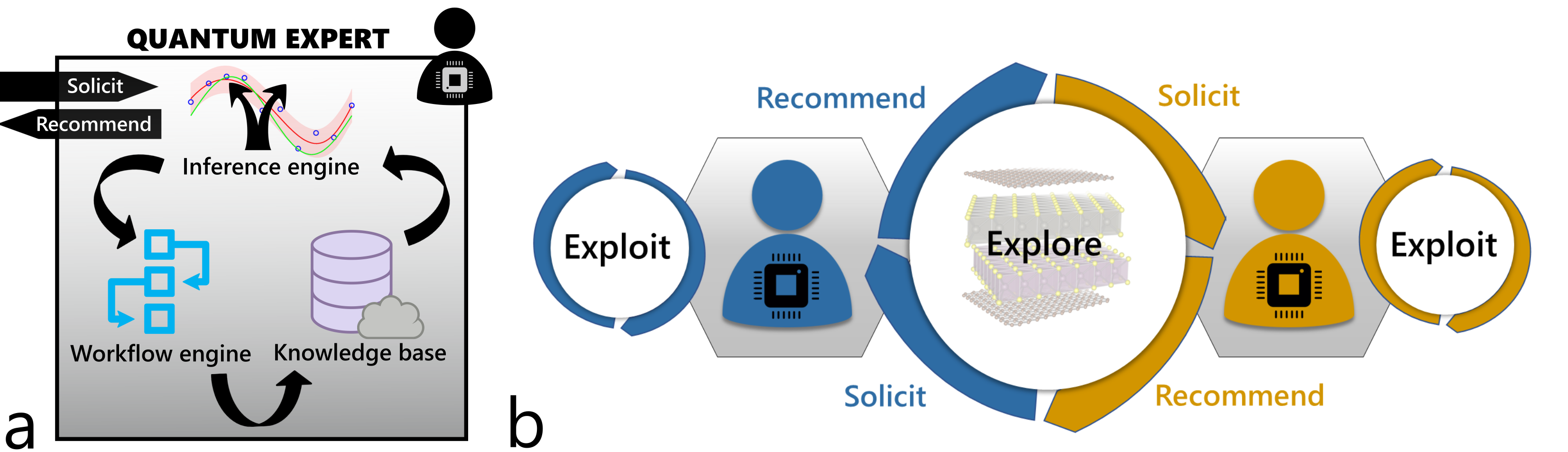}
\caption{a) Schematic of an {\em in silico Quantum Expert}, a unifying conceptual framework for guiding the collection and analysis of materials data. Key components are a planning inference engine, a workflow engine for measurements, and a database system. b) Modus operandi for closed-loop exploration of materials spaces using coupled instances of (a) as autonomous search agents. Each search agent is biased towards its own findings (exploitation) as well as promising regions of the search space discovered by the rest (exploration).}
\label{fig:quantumexpert}
\end{figure*}

Materials science and engineering have markedly benefited from the ability to identify working combinations of such technologies as high-performance computing, automation, and machine learning. Although a number of computational frameworks exist for leveraging high-performance computing to streamline materials calculations \citep{Jain2015FireWorks:,Pizzi2016AiiDA:}, they generally rely on extensions for optimization and decision-making in materials selection and design. Previously, we introduced the unifying conceptual framework of an {\em in silico Quantum Expert} (QE) for organizing materials calculations and directing the collection and analysis of data \citep{tritsaris_electronic_2020}, here implemented as a knowledge-based system with three key components (Figure~\ref{fig:quantumexpert}a): 
\begin{itemize}
    \item an inference engine that uses available materials information to plan measurements,
    \item a workflow engine for materials measurements to augment the knowledge base, and
    \item a database system as the storehouse of materials data and information (the knowledge base).
\end{itemize}
Lacking a closed-form solution to the problem of tailored design of moir{\'e} assemblies, we draw inspiration from the field of AI and use instances of the QE abstraction as information-sharing intelligent agents for simulation-based exploration of materials spaces (Figure~\ref{fig:quantumexpert}b). Each QE instance will:
\begin{enumerate}
    \item Generate a ranking of candidate materials according to how likely is their calculation to improve predictions of one or more properties across the entire materials space. For our application, the search agent effectively responds to the following question: ``{\em what layered assembly or assemblies should be calculated next to best improve predictions, given all previous outcomes?}''.
    \item Solicit rankings \blue{from connected search agents}, each generated based on their own scoring criteria while seeking to improve predictive models for the same or different properties. Pass own ranking \blue{to connected search agents}.
    \item Choose one or more materials for calculation and analysis while balancing exploitation, guided by its own ranking, and exploration, guided by external recommendations.
    \item Calculate the selected materials to augment the knowledge base. For our application, the search agent calculates materials by executing one or more predefined workflows in a high-throughput fashion.
    \item Return to step 1, and the cycle repeats.
\end{enumerate}
The simulation starts with a random sampling of the materials space, and finishes when uncertainty in predictions is reduced below a predetermined value or after a specified number of cycles. Our approach is positioned alongside \blue{global search methods that make minimal assumptions about the search problem (e.g., size of materials space, correlations among properties, derivative information), \citep{montoya_autonomous_2020,johannesson_combined_2002,wang_calypso_2012,Oganov2019} while exploiting uncertainty information to improve the search.} Therefore is particularly suitable for assessing the scope of novel materials concepts, the discovery of novel materials, or, for our application, explore the electronic tunability of complex moir{\'e} assemblies and broaden the materials space.

Briefly, the computational workflows (step `4' above) entail four main tasks \citep{tritsaris_electronic_2020}: 
\begin{enumerate}[label=\alph*.]
    \item Parse a string representing the layered assembly to be calculated \citep{tritsaris_lan_2020}.
    \item Generate and solve a TB model (see Section~\ref{sec:theoretical}).
    \item Analyze the calculated electronic energy levels, for example to construct measures of band flatness or extract such information as band gaps for electronic transitions.
    \item Insert the electronic band structure and any post-processing meta-data into the database. For our application, we rely on a document-based data model, as implemented in the document-oriented (NoSQL) database MongoDB.
\end{enumerate}
In summary, the above workflow transforms a string representing a layered assembly into descriptions of its electronic structure by means of setting up and solving a TB model. We have used established electronic structure methods and open source libraries for implementation. We provide specific implementation details in Section~\ref{sec:results}.

To be able to name, catalogue, and discuss the studied materials in an unambiguous fashion (see also step `a' above), we relied on a domain-generic {\em layered assemblies notation} corresponding to a theoretical materials concept of layer-by-layer robotic assembly of layered structures using a sequence of vertical stacking, rotation, or strain operations on individual layers \citep{tritsaris_lan_2020}. For example, the string `G/G@1.10’ describes a bilayer of graphene (often referred to in the literature as `TBG’) with counterclockwise twist angle $\theta = 1.10^\circ$, the string `G/G@1.50/G/G@1.50’ (or the shorter `2(G/G@1.50)’) describes an alternating twist quadruple-layer of graphene (`ATMG') with $\theta = 1.50^\circ$, and so on, with `G' the symbol for an extended graphene layer. Likewise, the string `m$_1$/m$_1$\#0.01' describes a 1D superlattice in which one layer of the material `m$_1$' is associated with $\Theta = 0.01$ (compressive strain of 1\%).

\section{Virtual Experiments} \label{sec:results}
\blue{We proceed to apply these concepts, models, and methods to study the electronic structure of various low-dimensional superlattices, and obtain quantitative insights into electronic tunability within these materials spaces.}

\subsection{Arbitrarily stacked multi-layer superlattices}
\label{sec:1d_assemblies}
To the best of our knowledge, there is no universal approach based on TB that is suitable for the calculation of arbitrarily stacked 2D superlattices. Compiling a library of TB parameters for selected materials is feasible, but nevertheless a daunting task. Furthermore, the calculation of materials more complicated than the prototypical graphene (which requires just two p$_z$ orbitals for its basis) becomes significantly more resource-demanding beyond two layers due to the large matrix size of the resulting Hamiltonian. 

As discussed in Section \ref{sec:theoretical}, 2D materials with twist-induced moir{\'e} patterns are directly comparable to the lattice-mismatch moir{\'e} patterns in 1D. We use six prototype 1D semiconducting materials as building blocks for 1D moir{\'e} assemblies (Figure~\ref{fig:1Dstructures}). By combining them, we compile an extended library of structures, a total of $\sim 200,000$ two-, three-, and four-layer assemblies, as required for reliably extracting broadly applicable physical insights. From a computational point of view, these have very low computational cost (a few seconds per calculations on a personal computer). We limit our investigation to a single value for $\Theta$ for each moir{\'e} assembly: each layer has either lattice constant $1$ or $1-\Theta$, preventing multiple incommensurations in three and four-layer structures \cite{Zhu2020}. For all material combinations, the interlayer separation is set to $3$ \AA{}, a choice whose primary purpose is to limit the maximum possible value of the interlayer coupling.  

\begin{figure*}
  \centering
  \includegraphics[width=\textwidth]{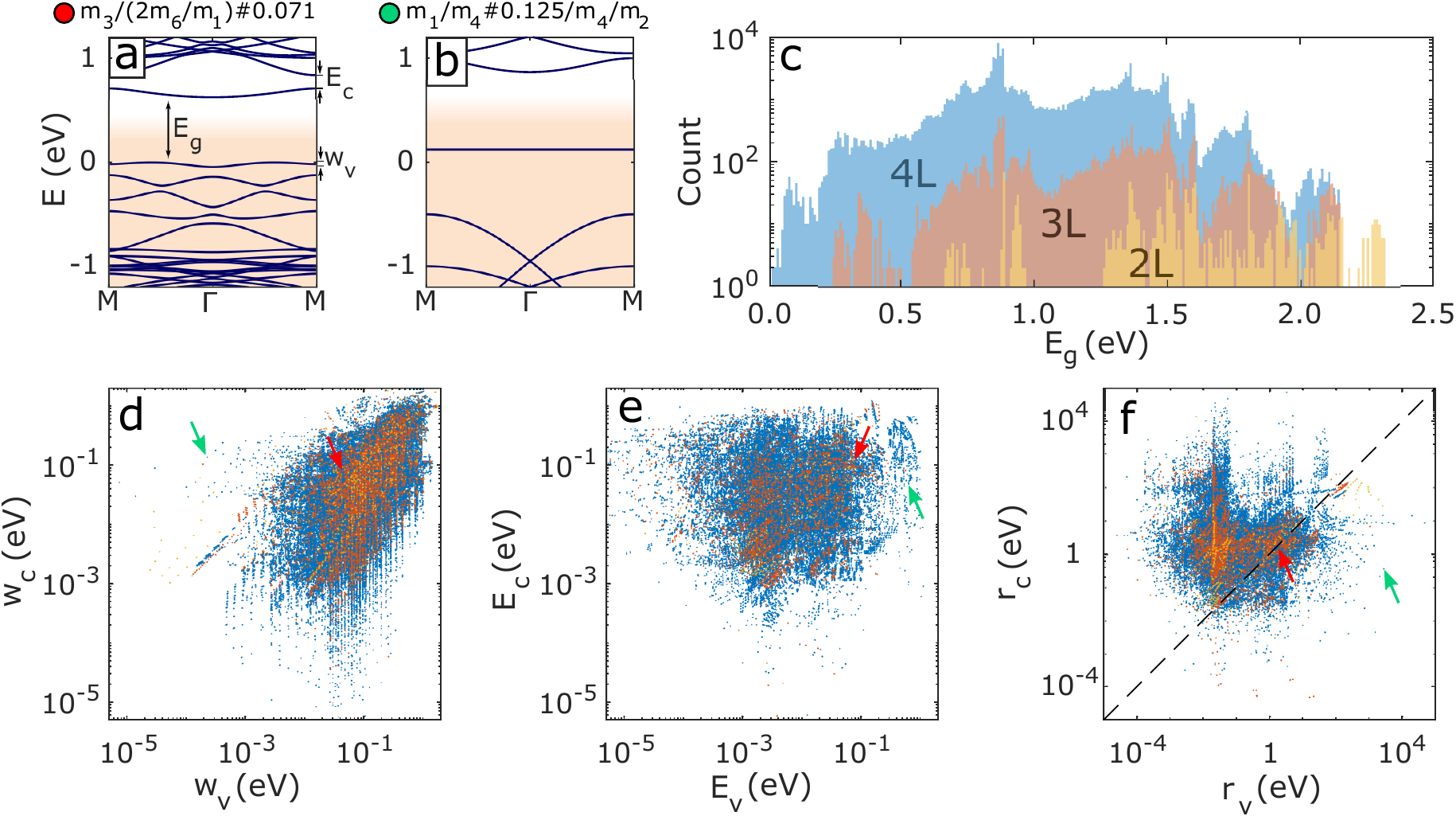}
\caption{Properties calculated by analysis of tight-binding band structures of model 1D moir{\'e} assemblies. (a,b) Example electronic band structures for 1D superlattices with (nearly-)flat electronic bands. The band gap, E$\rm{_g}$, valence bandwidth, w$\rm{_v}$, and conduction moir{\'e} gap, E$\rm{_c}$, are indicated in (a). (c) Number of moir{\'e} assemblies with a given band gap. Yellow, orange, and blue points and bars are associated with two- (2L), three- (3L), and four-layer (4L) assemblies, respectively. (d) Bandwidths of the valence and conduction (flat) bands, w$\rm{_v}$ and w$\rm{_c}$. (e) Moir{\'e} (direct) band gaps between the valence or conduction flat band and the next nearest band, E$\rm{_v}$ and E$\rm{_c}$. (f) Ratios between the bandwidths and moir{\'e} band gaps for the valence and conduction bands, r$\rm{_v}$ and r$\rm{_c}$. Bandwidths and band gaps are reported in eV. The colored marks next to the assembly formulas (a,b) match the arrows in (d-f), indicating their location on the plots.
}
\label{fig:1Dmaps}
\end{figure*}

With regards to characterization of the electronic structure, five direct properties and two derived properties are of interest:
\begin{itemize}
    \item the band gap for the moir{\'e} assembly, E$\rm{_g}$,
    \item the bandwidths of the valence and conduction (flat) bands, w$\rm{_v}$ and w$\rm{_c}$,
    \item the moir{\'e} (direct) band gaps between the valence or conduction band and the next nearest band, E$\rm{_v}$ and E$\rm{_c}$, and, derived from these,
    \item the ratios between the bandwidths and moir{\'e} band gaps for the valence and conduction bands, r$\rm{_v} = $ E$\rm{_v}/$w$\rm{_v}$ and r$\rm{_c} = $ E$\rm{_c}/$w$\rm{_c}$.
\end{itemize} 
The direct properties are indicated in Figure~\ref{fig:1Dmaps} for a selected 1D moir{\'e} assembly's band structure. Figure~\ref{fig:1Dmaps} also contains maps of all properties, as derived from the analysis of the TB band structure of each layered assembly, for all structures considered. In our calculations, materials without direct band gaps, either E$\rm{_g}$ or E$\rm{_{v,c}}$, yield negative values for the gaps, and although they are still tabulated they are omitted from the plots.

Two examples of moir{\'e} assemblies with flat bands, an indicator of interesting electronic behavior \citep{Carr2017Twistronics,Bistritzer2011Moire,Cao2018Unconventional}, are shown in Figure~\ref{fig:1Dmaps}a,b. The first of these, m$\rm{_3}/(2$m$\rm{_6}/$m$\rm{_1})\#0.071$ is a typical four-layer assembly: it has a band gap that is smaller than any of the original 1D monolayers and non-identical band flattening occurring between the valence and conduction edges. However, the second example, m$\rm{_1}/$m$\rm{_4}\#0.125/$m$\rm{_4}/$m$\rm{_2}$, is representative of an important group of outlier band structures: one band, which is significantly flatter than any other, seems to have moved into the middle of the band gap. Looking at Figure~\ref{fig:1Dstructures}, the likely origin of this effect is the proximity of m$\rm{_2}$'s valence band with the vacuum-level Fermi energy. Combined with a strong effective interlayer moir{\'e} potential at the twisted interfaces, the associated flat band separates from the surrounding electronic states. In this case, the definitions of $E\rm{_g}$ and $E\rm{_{v,c}}$ become interchangeable under adding or removing one electron from the moir{\'e} superlattice, and these large values of $E\rm{_{v,c}}$ explain many of the outlying large values of r$\rm_{{v,c}}$.

Focusing on trends, we identify a positive correlation between the bandwidths of the valence and conduction bands, spanning four orders of magnitude, with many stacks having (nearly-)flat electronic bands. It is also possible to identify structures with any combination of moir{\'e} band gaps, over four orders of magnitude as well. We find many structures associated with large ratios r$\rm{_v}$ and r$\rm{_c}$: such combination of properties would provide, for example, a clear optical signal by preventing additional unwanted peaks near the primary flat-band to flat-band transition. Showing large gaps but small bandwidths is also important for correlated phases. For example, if the effective Hubbard parameter for the flat band is larger than the moir{\'e} gap, then candidate Mott insulating states will necessarily mix the next set of moir{\'e} bands, possibly preventing a correlated ground state.

\blue{One exceedingly promising result, which is expected to apply to 2D materials as well,} is that we find layered assemblies with almost any electronic band gap for electronic transitions, direct or indirect, between a practically continuous range (between 0 and 2.0~eV) from combinations of only six unique materials. The origin of this phenomena can be understood by looking at the layer-scaling of the band-coverage in Fig. \ref{fig:1Dmaps}c. The gaps of the two-layer structures do not densely cover any range, while the three and four-layer structures become progressively denser in their coverage of band gap energies. Each constituent layered material has its own band gap, but when it is combined with another material the interlayer coupling causes hybridization at the conduction and valence band edges, generally reducing the band gap of the layered assembly. This gives every unique assembly its own ``composite'' band gap, even in the absence of a moir{\'e} pattern. Strain (or the twist angle in 2D) then provides fine-tuning of the effective interlayer couplings strength, and generally causes gaps to widen (this effect will be studied in detail in the following section for MoS$_2$). As the number of layers in the assembly increases, the number of unique assemblies with their own unique band gap grows quickly: with just four layers, complete coverage of the spectrum is possible.

Overall, Figure~\ref{fig:1Dmaps} illustrates how the stacking of two- (yellow points and bars), three- (orange), or four- (blue) layers results in properties with an increasingly wider and smoother distribution of values as the number of layers increases, with four layers sufficient to cover a large range in a continuous fashion. Therefore, by enumeration we infer that it is possible to design moir{\'e} assemblies with finely tuned electronic properties, even with a small number of materials and layers. Such complete control has clear technological benefits for the development of tailored layered materials for photovoltaics, energy storage, quantum information, and other technologies. 

Nevertheless, due to the very large number of structures, performing an exhaustive search of such large materials spaces is an impractical task, especially when objective evaluations are expected to be limited, due to finite resources (computational or otherwise). Masubuchi {\em et al.} \citep{masubuchi_autonomous_2018} introduced an approach to the high-throughput robotic assembly of 2D crystals for designer multi-layer assemblies. Consider a simulator of such automated system that relies on virtual high-throughput measurements (calculations) of layered materials assembled from a finite set of single-layer building blocks. We instantiate the QE abstraction for this simulation, using multiple instances as information-sharing search agents that follow the protocol of Section~\ref{sec:computational}, with the aim to obtain a reliable surrogate model of all properties of interest while drastically reducing needed calculations. Specifically, each QE instance will:
\begin{enumerate}
    \item \blue{Generate a ranking for 1D moir{\'e} assemblies for one property of interest, based on the sampling score $\alpha$. For a not-yet-calculated structure S, the sampling score is calculated as:
    $$\alpha[{\rm S}] = \frac{1}{2} \sigma[{\rm S}] + \frac{1}{2} \frac{\partial \mu[{\rm S}]}{\partial \Theta}.$$
    The first of the two terms (each normalized to the range [0,1]) aims to improve overall prediction accuracy by directing the sampling towards materials subspaces associated with relatively high uncertainty in the predicted property, whereas the second term aims to improve predictions within each configuration with respect to the single most important ``local'' feature, the parameter $\Theta$. In this scheme, the higher $\alpha[S]$ is for a structure S, the more likely this structure will be chosen for calculation. For predictions, a fully-connected neural network is used: by applying the same input to the neural network many times, an empirical distribution over the property is deduced, which is used to obtain a mean value, $\mu$, and the corresponding standard deviation, $\sigma$. The neural network comprises two hidden layers of 100 and 20 nodes. A dropout layer that randomly sets inputs from the first hidden layer to 0 with a 50\% probability is introduced to avoid overfitting during training and for quantifying uncertainty in predictions. Feature vectors are constructed using a distinct one-hot encoding of each assembly, augmented with a list describing lattice mismatch (0 or $\Theta$).}
    \item Solicit rankings \blue{from connected search agents} tasked to improve estimations of other properties. Pass own ordered list of candidates \blue{to connected search agents} as a recommendation.
    \item Choose for calculation and analysis either structure S$_i$ if
    $$u_{ij} < 1 - \frac{\beta e^{-\alpha[\rm{S}_i]}}{\beta e^{-\alpha[\rm{S}_i]}+(1-\beta) e^{-\alpha[\rm{S}_j]}},$$
    or structure S$_j$ otherwise, with random number $u_{ij} \in [0,1]$. The parameter $\beta$ controls the balance between exploitation ($\beta$~=~0) and exploration ($\beta$~=~1). The indices describe the internal (`i') and solicited (`j') rankings, from which pairs of candidate materials (S$_i$, S$_j$) are drawn. For instance, for a pair with $\alpha[S_i]-\alpha[S_j]=0.5$ and $\beta=1/2$, there is a 38\% chance that the search agent will be forced to choose S$_j$, despite its preference according to its own ranking. \blue{For $\alpha[S_i]=\alpha[S_j]$, the probability is $1-\beta$, {\em i.e.,} 50\% chance for $\beta=1/2$}. For our application, a batch of twenty-five moir{\'e} assemblies are identified and scheduled at each cycle for calculation (less than 1/10,000th of all possible combinations).
    \item Calculate and analyze the band structure of the selected structures using TB.
    \item Return to step 1, and the cycle repeats.
\end{enumerate}
The simulation starts with a random sampling of the materials space, and finishes after a specified number of cycles. \blue{For our application, each search agent is connected to one other for receiving a solicitation and a different one for passing recommendations, both of which are assigned at random at the start of each cycle}. This setup is reasonable, given that changes in the band structure near the Fermi level which induce change in one property may be associated with change in other properties as well. Likewise, the choice of neural networks as predictive models is not unique, but we leverage them as universal approximators and use the same generic architecture for all search agents without assuming {\em a priori} knowledge about the materials space or properties ({\em e.g.,} correlations between them). To the casual materials scientist, the network of QE instances functions as a virtual research lab with its devices working continuously and interactively to refine a surrogate model of structure-property relationships within the space of the 1D moir{\'e} assemblies. 

For $\beta = 1/2$, after only $3\%$ of the materials space has been sampled, the network of QE instances has collectively learnt to predict the five electronic structure properties of interest across the entire space within a typical mean squared error (MSE) of less than 0.01~eV$^2$, with an average coefficient of determination $R^2 = 0.63$. The good performance can be attributed in part to the smooth evolution of the electronic bands with respect to the parameter $\Theta$ (see also Figure~\ref{fig:mos2}).

We also performed simulations in the limiting cases $\beta = \{0,1\}$, with $\beta~=~0$ forcing each search agent into closed-loop feedback, and $\beta~=~1$ forcing each search agent to rely solely on external recommendations. In both cases, the predictive capabilities of the search agents deteriorated, with 15\% and 100\% higher MSE (20\% and 65\% smaller $R^2$) for $\beta~=~0$ and $\beta~=~1$, respectively, demonstrating the benefit of carefully balancing exploration with exploitation. \blue{Furthermore, for a setup in which how the search agents are connected to one another for receiving solicitations and passing recommendations do not change, the number of required cycles increased by more than 25\% to achieve about the same MSE and $R^2$ in predictions.} 

Therefore, we arrive at the conclusion that it is possible to target any value for important electronic structure properties through careful choice of the constituent materials and their stacking geometry, and that the identification of such a structure should be computationally tractable even for very large material spaces. \blue{We expect these findings (pertaining to multiple materials, layers, and properties) will carry over from 1D superlattices to the case of 2D layered materials. We reiterate that the band flattening and corresponding correlated phases in twisted 2D materials, or the electronic tunability seen in the 1D systems here, are not specific to the dimensionality of the crystal. Rather, both of these phenomena are caused by the effective superlattice generated by relative strain between the stacked layers. The tunable length-scale of such superlattices allows for the engineering of electronic modes across a wide range of energy scales.}

\subsection{A bilayer of molybdenum disulfide}
\blue{In this and the next section we investigate two key ingredients of findings pertaining to 1D moir{\'e} structures in realistic 2D systems, namely, band tunability through the twist angle and the degree of diversification of interface couplings in multi-layer assemblies.}
 
For twisted 2D semiconductors, a typical model system is the twisted bilayer MoS$_2$ (MoS$_2$/MoS$_2$@$\theta$). We will focus on a key property that mirrors the $\Theta$-tunability associated with the 1D assemblies, the twist angle(s) at which the dispersion of one or more electronic bands near the band edges significantly reduces. Alongside layered graphene, layered 2D semiconductors like MoS$_2$ exhibit non-trivial physical behavior that depends sensitively on the number and relative orientation of the constituent layers \citep{huang_low-frequency_2016,yeh_direct_2016,Carr2017Twistronics, Zhai2020}. Another member of this technologically important class of materials, tungsten diselenide (WSe$_2$), was recently observed to host correlated insulating states in its two-layer assembly \citep{Wang2020,Zhang2020}, joining graphene as a strongly-correlated moir{\'e} material.

\begin{figure*}
  \centering
  \includegraphics[width=\textwidth]{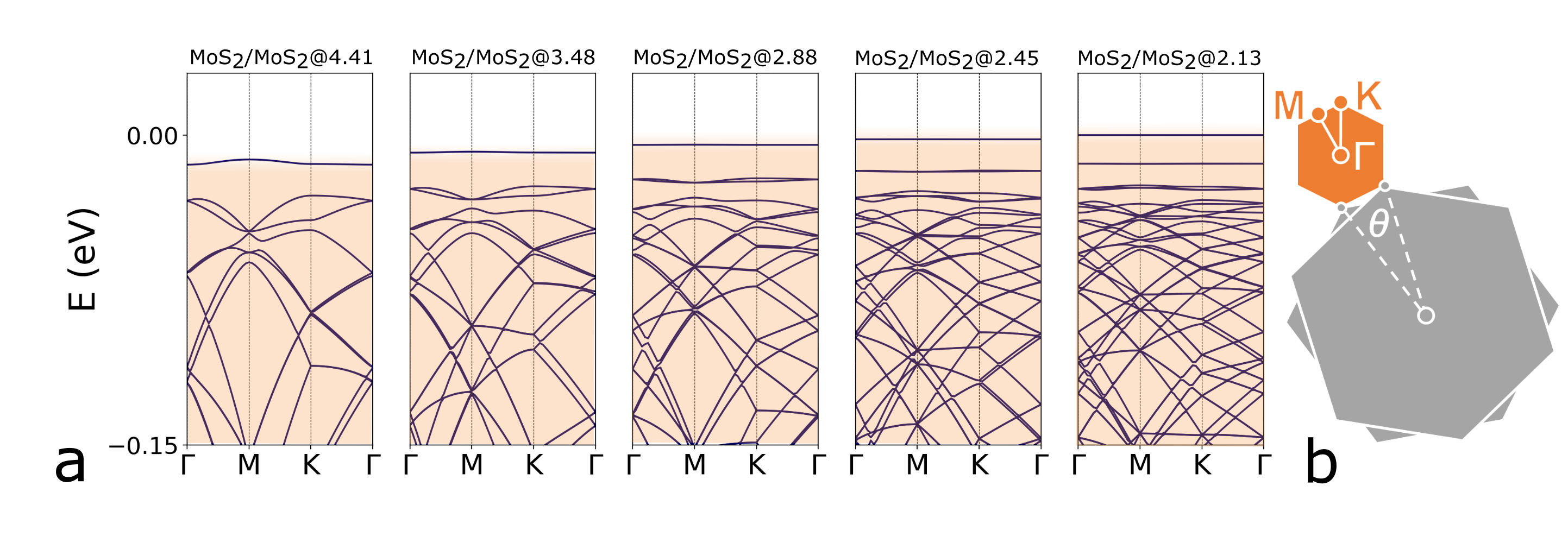}
\caption{a) Low-energy electronic band structure (valence bands only) for twisted bilayers of MoS$_2$ at relative small twist angles, obtained using an {\em ab initio} tight-binding model. The shaded area marks occupied electronic states. b) Brillouin zone of isolated layers (grey) and twisted superlattice (orange) with high-symmetry points $\Gamma$, K, and M.}
\label{fig:mos2}
\end{figure*}

We used atomistic structural models of superlattices consisting of two sheets of MoS$_2$ in AA stacking with an in-plane lattice constant of 3.18 \AA{}, and interlayer distance fixed at 6.145 \AA{}, to calculate all commensurate superlattices between `MoS$_2$/MoS$_2$@21.79' and MoS$_2$/MoS$_2$@1.02. A sequence of thirty-two commensurate bilayers of MoS$_2$ were calculated (Figure~\ref{fig:mos2}).

The low-energy band structures for large twist angles resemble that of an isolated MoS$_2$ layer, with little effective interaction between the two layers. As the twist angle decreases, strong interlayer hybridization leads to smooth splitting and flattening of electronic bands near the valence band maximum. Figure~\ref{fig:mos2} shows TB band structures (valence bands only) of MoS$_2$ bilayers near the critical $\theta = 2.9^\circ$, and the corresponding Brillouin zone of isolated layers (grey), and twisted superlattice (orange). At $\theta = 2.1^\circ$, an electronic band with nearly zero bandwidth coexists with a pair of almost degenerate flat bands at slightly lower energy, while the next three bands have begun to flatten out. This sequence of flat bands follows the degeneracies of the 2D quantum harmonic oscillator, as predicted by continuum models which consider the moir{\'e} patterns as a periodic network of potential wells for the bilayers' electrons \citep{Carr2020duality,angeli2020gammavalley,Zhai2020}. Such separation of sets of bands is often desirable in an experimental setup because it enables controlled modulation of transport properties via the application of external fields. Nonetheless, since the electronic bands evolve smoothly with respect to the twist angle, the small parameter space in the simple case of the twisted bilayer of MoS$_2$ leaves little room for electronic tunability: more layers are needed to obtain a larger range of accessible electronic behavior and properties.

\subsection{Twisted multi-layer graphene assemblies}
Graphene-based moir{\'e} assemblies, regarded by many scientists and technologists as the quintessence of 2D layered materials, continue to be investigated by transport measurements, theoretical modeling, and computation \citep{Cao2018Unconventional,xie_spectroscopic_2019,choi_electronic_2019,tritsaris_electronic_2020,luo_situ_2020}. Driven by the hypothesis that increasingly complex 2D moir{\'e} assemblies enable more precise control of electronic properties and functionality, we next extend our investigation to a much larger set of structures than previously, {\em i.e.,} moir{\'e} superlattices of two to six, AB-stacked, sheets of graphene. We specifically examine the effect of an assembly's complexity (size, stacking order, and type of interfaces) on the twist angle(s) at which the dispersion of one or more electronic bands near the Fermi level significantly reduces.

We limit our investigation to moir{\'e} assemblies with a single twist angle for any rotated layers, including a twisted graphene monolayer encapsulated by two or more additional sheets such as 2G/G@$\theta$/2G, alternating relative twist angle assemblies such as 2(G/G@$\theta$)/G, and a twisted trilayer of graphene on another trilayer, 3G/3G@$\theta$. We employed rigid atomic sheets of graphene, with an in-plane lattice constant of 2.47~\AA{}, and interlayer distance fixed at 3.35~\AA{}, to obtain single-particle band structures for 640 unique commensurate superlattices of twisted graphene sheets in total. 

\begin{figure*}
  \centering
  \includegraphics[width=\textwidth]{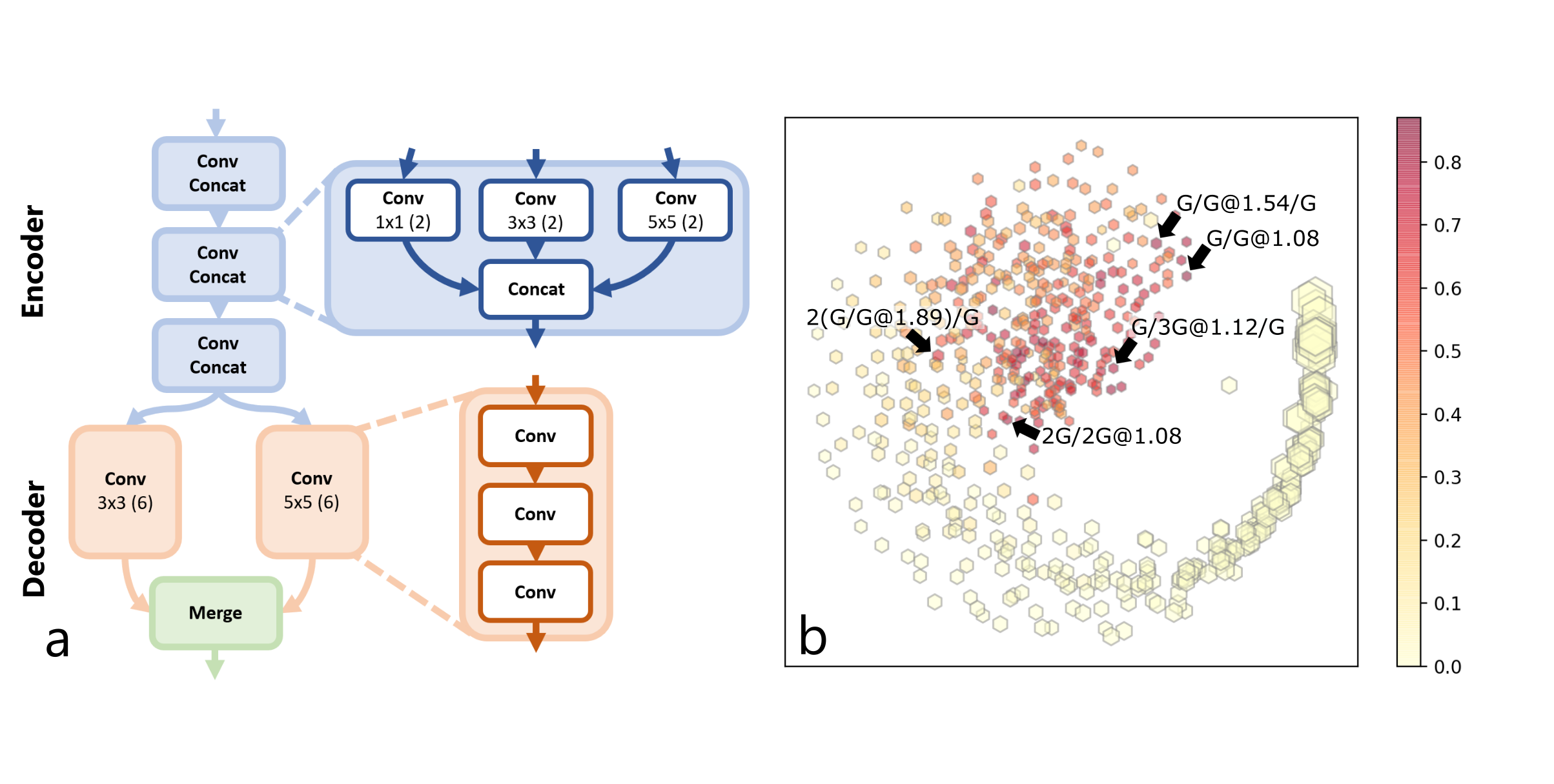}
\caption{(a) Convolutional neural network with an autoencoder (encoder-decoder) architecture for compressing the calculated tight-binding band structures of multi-layer graphene. (b) Visualization of the resulting 2D embedding. Points in the right panel are color-coded with respect to the likelihood of interesting electronic behavior (white for small; red for high), and the symbol size is proportional to the twist angle. The black arrows indicate where (selected) layered assemblies with low-dispersion bands can be found (see also Figure~\ref{fig:56L}).}
\label{fig:atlas}
\end{figure*}

To identify patterns across the entire materials space, we generate a single 2D visualization of the database in a state that contains all calculated band structures. This is achieved with a convolutional neural network that implements an autoencoder (encoder-decoder) architecture, shown in Figure~\ref{fig:atlas}a, treating each band structure as an image. Convolutional networks are central to state-of-the-art approaches in computer vision and image processing for detecting important features without any supervision \citep{szegedy_rethinking_2016,noauthor_efficient_2007}. For our application, this approach has the advantage of being immune to artificial band crossings in the calculated TB band structures, which might complicate less sophisticated data analysis using simple descriptors, while retaining information about electronic bands further from the Fermi level.

Each of the three modules in the encoder compresses its input by performing convolutions with two (2) filters of three different sizes (1$\times$1, 3$\times$3, or 5$\times$5; strides of 2$\times$2). The outputs are concatenated into a single output vector forming the input of the next module. The decoder learns to reconstruct the original input from the reduced representation; it comprises two modules, each performing three convolutions using six (6) filters of one size (3$\times$3, or 5$\times$5; strides of 2$\times$2). The outputs of these two modules are merged, for example by computing the element-wise maximum. \blue{Most importantly, the use of different types of filters aims to handle features at multiple scales better with parallel gains in performance, as in Inception-motivated neural network architectures for image classification \citep{szegedy_rethinking_2016}}. We implemented this architecture using Keras (version 2.3.1), a deep learning API written in Python \citep{chollet2015keras}, running on top of the machine learning platform TensorFlow (version 2.1.0) \citep{abadi2016tensorflow}. 

After training, we use the convolutional autoencoder to compress each band structure in a window of 0.30~eV centered at the Fermi level, fed to the network as 40$\times$40 binary images, into 150-dimensional vectors. Exploiting symmetries in the electronic band structures, we augment the data set by mirror and flip operations to improve the performance of the model. Although exhaustive hyperparameter optimization is beyond the scope of this study, this architecture was found to outperform in terms of accuracy fully-connected architectures with similar number of trainable parameters ($\sim 9000$) or number of layers (92\%, using a 80/20 training/validation split).

The vectorized representations are shown in Figure~\ref{fig:atlas}b as a 2D embedding, prepared using multidimensional scaling (MDS), as implemented in the scikit-learn package (version 0.20.3), a machine learning library for Python \citep{Pedregosa2011Scikit-learn:}, on the basis of a cosine distance metric for evaluating pairwise distances between the 150-element vectors (another reasonable choice would have been the t-distributed stochastic neighbor embedding method, t-SNE). The transformation favors a clustering in which the smaller the distance between any two points in the resulting abstract Cartesian space is, the more similar the corresponding band structures are. \blue{Overall, Figure~\ref{fig:atlas}b provides a concise picture of relationships across the entire materials database that is not readily observable by means of enumeration.}: points are color-coded with respect to the likelihood of interesting electronic behavior (white for small; red for high), using the flatness of electronic bands near the Fermi level in the TB band structures as a descriptor (for more details about this separate characterization, also based on image analysis, see Tritsaris {\em et al.} \citep{tritsaris_electronic_2020}) 

The primary ordering of the generated 2D embedding roughly follows $\theta$, even though such information was not explicitly used to create it. \blue{Many layered assemblies at relatively small twist angles (corresponding to stronger relative interlayer hybridization) exhibit electronic bands of low dispersion}. This observation can be rationalized by noting the increased degrees of freedom thicker superlattices have, {\em i.e.,} possible stacking and rotation combinations, \blue{which readily carries from the case of 1D moir\'{e} assemblies,} corroborating our working hypothesis. Equivalently, simpler layered assemblies can be thought of as building blocks whose electronic structure is perturbed when embedded in more complex assemblies. \blue{The ``building block'' nature of the moir\'{e} interfaces can also be seen when examining the specific TB electronic band structures. Those with demonstrably low-dispersion bands (Figure~\ref{fig:56L}) are highlighted below for the five- and six-layer moir{\'e} assemblies. As structures with more than four layers are uncommon in the current literature on twistronic graphene, many of these assemblies are modeled here for the first time. In addition, these calculations provide reference for experimental measurements such as accurate estimations of magic angles}:

\begin{figure*}
  \centering
  \includegraphics[width=\textwidth]{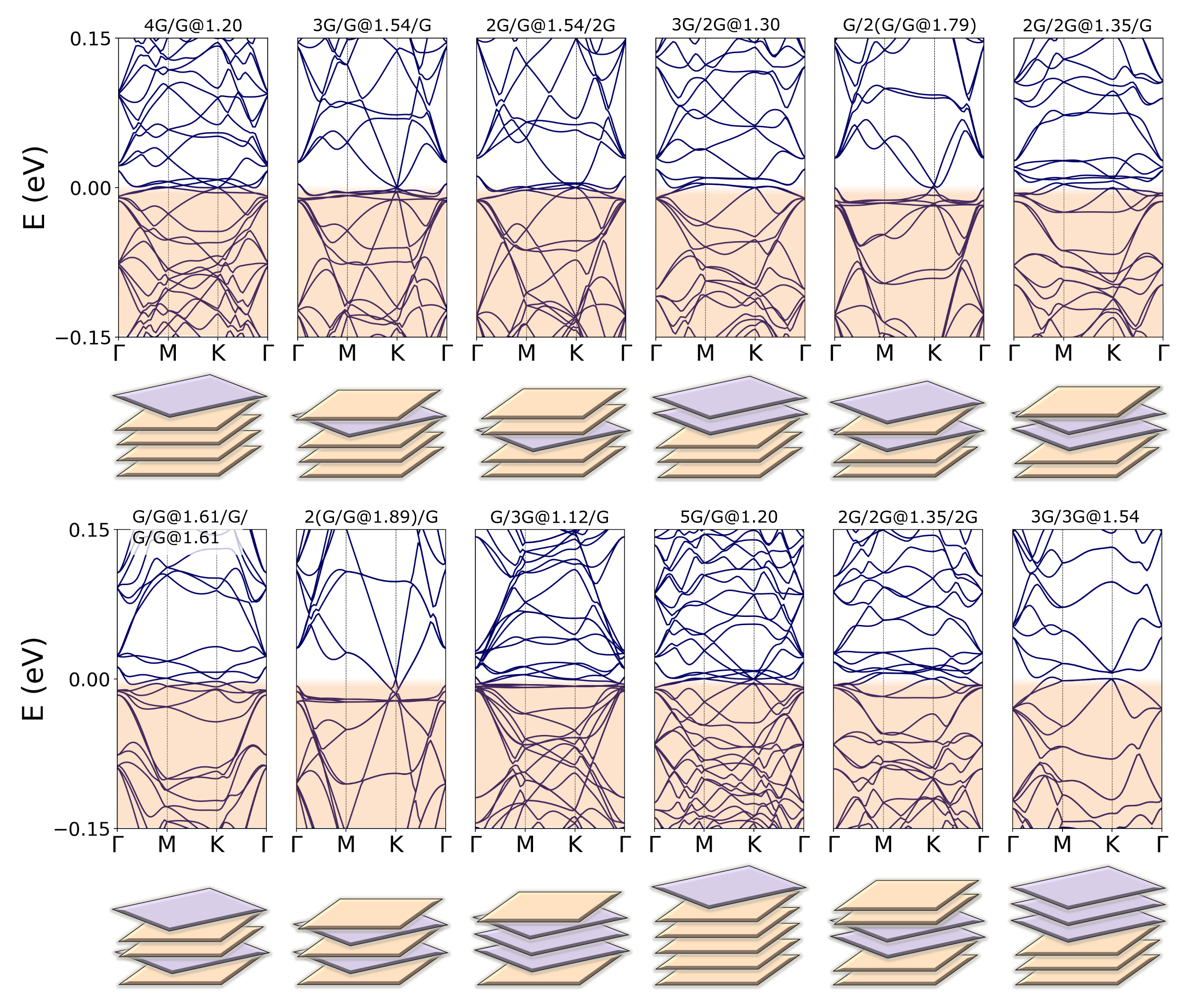}
\caption{Low-energy electronic band structure for twisted five- and six-layer assemblies of graphene with low-dispersion or almost-flat electronic bands, obtained using an {\em ab initio} tight-binding model. The shaded area marks occupied electronic states.}
\label{fig:56L}
\end{figure*}

4G/G@$\theta$. The twisted monolayer on a quadruple-layer of graphene has a magic angle near $\theta = 1.2^\circ$. At the supercell K-point a cone-like feature remains but has parabolic dispersion. This parabolic dispersion of this feature is indicative of the presence of bulk Bernal-stacked graphitic band structure, {\em e.g.,} there is a nearly undisturbed ``AB'' bilayer graphene band structure present.

3G/G@$\theta$/G. This configuration exhibits bands with low dispersion near $\theta = 1.5^\circ$. An outstanding feature is the coexistence of a Dirac cone that resembles that of single-layer graphene. This symmetry-protected cone is also observed in the ``sandwich'' configuration G/G@$\theta$/G \citep{carr_ultraheavy_2020}, and the parabolic ``AB'' bands that approach near the Fermi level resemble the case of 2G/G@$\theta$/G \citep{tritsaris_electronic_2020}.

2G/G@$\theta$/2G. Similar to the previous configuration, however the parabolic ``AB'' bands are now symmetry-protected and thus are not gapped at the K-point.

3G/2G@$\theta$. We identify a region of nearly-flat bands at the Fermi level near $\theta = 1.3^\circ$ for a twisted bilayer on a trilayer. Compared to a twisted double bilayer 2G/2G@$\theta$, this configuration also exhibits parabolic bands near the Fermi level \citep{tritsaris_electronic_2020}, but the broken mirror symmetry has opened up a sizable gap at the K-point for both the flat and parabolic bands.

G/2(G/G@$\theta$). This configuration resembles the class of layered assemblies of graphene with alternating relative twist angle, showing a pair of Dirac bands at the K-point and flat bands at a relatively higher magic angle approaching two times this of a twisted bilayer as the number of layers increases \citep{Khalaf2019Magic}. In this case too, the z-mirror symmetry breaking caused by the additional layer on the bottom of the stack has caused avoided band crossings near the K-point.

2G/2G@$\theta$/G. This layered assembly mostly resembles twisted double bilayer 2G/2G@$\theta$ \citep{Lee2019Theory}, however the additional top layer has made some important changes: aside from the usual gap openings at the K-point, there are also new flat bands away from the Fermi level.

G/G@$\theta$/G/G/G@$\theta$. Electronic bands with low dispersion are identified near $\theta = 1.6^\circ$. This combination, owing to symmetry, shows similar features to the encapsulated bilayer G/2G@$\theta$/G, and no well-defined Dirac cones.

2(G/G@$\theta$)/G. The layered assembly of graphene with alternating relative twist angle exhibits magic angle flat bands near $\theta = 1.9^\circ$. This magic angle is higher than the related three- and four-layer configurations G/G@$\theta$/G and 2(G/G@$\theta$/G), in agreement with the theoretical work of Khalaf {\em et al.} \citep{Khalaf2019Magic}, based on model Hamiltonians.

G/3G@$\theta$/G. For a twisted graphene trilayer encapsulated by another two, well-defined and low-dispersion bands form near $\theta = 1.1^\circ$. The TB electronic band structure exhibits mixed character between that of a twisted bilayer G/G@$\theta$, and a twisted monolayer on a bilayer 2G/G@$\theta$. This structure can be thought of as two copies of the twisted bilayer glued together weakly by a middle layer, causing more band dispersion but keeping the same overall structure.

5G/G@$\theta$. As in the case of all other twisted monolayers on graphite-like substrates, a twisted monolayer on a five-layer substrate shows strong graphitic character with parabolic and Dirac-like bands at the high-symmetry K-point. As the number of layers on the bulk-like side increases, the band structure will show increasing numbers of parabolic bands at the K-point, and eventually the band structure will be better described by including an additional $k_z$ momentum ({\em e.g.,} a transition from a 2D to 3D band structure).

2G/2G@$\theta$/2G. This is an alternating double-bilayer graphene assembly, and is the sandwich generalization of the twisted double bilayer. Unlike G/G@$\theta$/G, no large enhancement in the magic angle is observed, and overall the band structure resembles this of the double bilayer. This is because making the constituent elements of the encapsulated double bilayers, instead of monolayers, means each successive moir{\'e} interface is only perturbatively coupled due to the relatively weak interlayer coupling between aligned layers (roughly a factor of 1/10 smaller than the in-plane coupling).

3G/3G@$\theta$. The low-energy bands in twisted trilayer graphene show the largest bandwidth of the six-layer assemblies studied here.
Two bulk graphite slabs twisted on top of one another will still host flat bands at the moir{\'e} interface, but the localized modes tend to be more dispersive than those of twisted bilayer graphene G/G@$\theta$, or of a twisted monolayer on bulk \citep{Cea2019Twists}.
However, reports of superconducting behavior in pyrolytic graphite attributed to twisted interfaces precede those of G/G@$\theta$ \citep{Scheike2012,Ballestar2015}.
Therefore, although the bands of this structure are not as flat as some of the other six-layer candidates, it should not be discounted as a viable platform for studying strongly correlated phases.

A discussion of trends across two-, three-, and four-layer assemblies is provided in earlier work of ours \citep{tritsaris_electronic_2020}. Briefly, we previously identified the following magic-angle structures: G/G@1.1, 2G/G@1.1, G/G@1.5/G, 3G/G@1.1, 2G/G@1.5/G, 2G/G@1.1, G/2G@1.7/G, and 2(G/G@1.7). In summary, these findings indicate that increasingly thicker 2D layered assemblies enable meaningful tunability of electronic properties, even when only one material and a relatively small number of layers are considered. \blue{Moreover, we anticipate the high-throughput computational approach demonstrated previously for 1D moir{\'e} assemblies to be as useful for large libraries of arbitrarily stacked 2D multi-layer superlattices, once physics-based models become available for their high-throughput  calculation.} 

\section{Conclusions and outlook} \label{sec:conclusions}
We introduced concepts, models, and methods for the automated planning and execution of (virtual) materials measurements and used them to explore spaces of model moir{\'e} assemblies in 1D, twisted two-layer MoS$_2$, and twisted multi-layer graphene. Tight-binding band structures were obtained, and examined with the primary aim to examine the degree that increasingly complex layered assemblies, in terms of variety of constituent single layers, number of layers, and type of interfaces, enable finer control of electronic properties, with implications for the design of active materials for devices.

\blue{A very important finding is that even a small set of materials is sufficient for the engineering of tailored moir{\'e} assemblies, at least within a particular application domain. Effectively, the problem of the (computational) design of moir{\'e} assemblies is reduced to the problem of identifying minimal sets of relevant single-layer building blocks, tailored for the intended application. Moreover, the combination of twisted 2D layers with intercalation or strain can provide the means for spatial control over materials properties and an additional knob with which to tune moir{\'e} physics \citep{larson_effects_2020,tritsaris_first-principles_2019}.}

Our study provides specific insights into the electronic structure of layered MoS$_2$ and graphene, and serves as a useful reference for future study of these materials. Calculations of two-layer assemblies of MoS$_2$ reveal multiple flat bands separated by tens of meV below 3$^\circ$ twist, in good qualitative agreement with previous modeling approaches of bilayer transition metal dichalcogenides \citep{Naik2018,Carr2020duality,angeli2020gammavalley,Zhai2020}. The diverse array of band structures for five- and six-layer graphene assemblies show that specific types of linear or parabolic band crossings within a magic-angle flat band manifold can be engineered by careful combination of simpler graphene stacks. Here, we do not account for the effect of structural relaxation, which can modify the band structures significantly, especially at lower twist angles \citep{Carr2019Exact}. For that reason, we did not consider 2D layered assemblies with very small rotations ($\theta < 1^\circ$). 

\blue{Our computational framework, as described in \ref{sec:computational}, is completely general} and should be applicable to the automated discovery and design of other materials as well, describing in a uniform way such workflows \citep{masubuchi_autonomous_2018, montoya_autonomous_2020, bassman_active_2018}. Depending on the details of the search problem, various \blue{ implementations} can be pursued, including using multiple search agents for the prediction of a single property of interest, a single agent for the prediction of multiple properties, on-the-fly adjustment of the balance between exploration and exploitation (the parameter $\beta$), and so on. 

Combining virtual with physical experiments in a systematic fashion is expected to also be beneficial for tailored design of layered assemblies: consider a network of information-sharing virtual and physical devices for screening of moir{\'e} assemblies in a closed prediction/verification loop. We hold the view that, presently, this is the shortest route towards accelerating the transition from the formulation of a novel materials concept to the development of a working prototype device with tailored functionality. On the one hand, despite the fact that high-throughput experimentation is an established method for exploring materials spaces, the combinatorial nature of the problem of designing twisted layered materials renders the approach virtually impractical. On the other hand, even a scalable, high-throughput computational approach as ours may still require thousands of calculations, which can be resource-intensive for realistic structural models. For example, commensurate structural models of bilayers of MoS$_2$ with twist angles near the critical angle (Figure~\ref{fig:mos2}) comprise $\sim 10,000$ atoms, although these calculations remain tractable, especially when a directed search of the space is conducted using predetermined targets for the properties of interest. 

To conclude, the conceptual frameworks, theoretical models, and computational methods we have introduced in this and recent related work \citep{tritsaris_electronic_2020,tritsaris_lan_2020,Carr2020review,Carr2020duality} collectively constitute building blocks for a solid foundation of twisted multi-layer assemblies as a distinct field of inquiry at the interface of materials theory and computational science within the broader knowledge domain of complex surfaces and interfaces.

\section*{Acknowledgements}
The authors would like to acknowledge Efthimios Kaxiras and Pavlos Protopapas at Harvard University for stimulating discussions. 

Electronic structure calculations were performed on the Extreme Science and Engineering Discovery Environment (XSEDE), which is supported by NSF grant No. ACI-1548562. Machine learning models were trained on the Cannon cluster, supported by the FAS Division of Science Research Computing Group at Harvard University. We relied on computational resources of the National Energy Research Scientific Computing Center (NERSC), a DOE facility operated under Contract No. DE-AC02-05CH11231, for the provisioning of databases supporting the computational work.

This work was supported in part by DOE Office of Science (Basic Energy Sciences; BES) under Award No. DE-SC0019300, by NSF grant No. OIA-1921199 and No. DMR-1231319 (Science and Technology Center on Integrated Quantum Materials, CIQM), and by the S{\~a}o Paulo Research Foundation (FAPESP) under grant No. 17/18139-6.

%\section*{Data availability}
%he data that support the findings of this study are openly available on the Materials Cloud at https://doi.org/10.24435/materialscloud:7e-pc.

\section*{References}

\bibliography{References}

\end{document}